\documentclass[11pt,twoside,letterpaper]{article}
\usepackage{times,fancyhdr}
\usepackage{graphicx}
\newcommand {\beq}{\begin{equation}}
\newcommand {\eeq}{\end{equation}}
\newcommand {\beqa}{\begin{eqnarray}}
\newcommand {\eeqa}{\end{eqnarray}}         %Equation version
\newcommand {\beqs}{\begin{eqnarray*}}
\newcommand {\eeqs}{\end{eqnarray*}}
\newcommand {\bds}{\begin{displaymath}}
\newcommand {\eds}{\end{displaymath}}
\newcommand {\n}{\nonumber\\}

\newcommand{\no}{\noindent}
\newcommand {\bebb}{\begin{thebibliography}}
\newcommand {\eebb}{\end{thebibliography}}      %Reference version
\newcommand {\bbit}{\bibitem}
\def\a{\alpha}

\def\G{\Gamma}

\def\lt{\left}
\def\rt{\right}
\def\rtarr{\rightarrow}

\def\dg{\dagger}

\def\journal#1&#2(#3){\unskip, \sl #1\ \bf #2 \rm(19#3) }
\def\andjournal#1&#2(#3){\sl #1~\bf #2 \rm (19#3) }

\def\npb#1#2#3{\textit{Nucl. Phys}. {\bf B#1}, (#2) #3}

\def\plb#1#2#3{\textit{Phys. Lett}. {\bf B#1}, (#2) #3}

\def\jmp#1#2#3{\textit{J. Math. Phys}. {\bf #1}, (#2) #3}

\begin{document}
\title{\bfseries\scshape{Super Coherent States, Boson-Fermion Realizations and 
Representations of Superalgebras}}
\author{\bfseries\itshape{Yao-Zhong Zhang}\\
Department of Mathematics, University of Queensland,\\ Brisbane, Qld 4072, 
Australia}

\date{}
\maketitle
\thispagestyle{empty}
\setcounter{page}{1}

\pagestyle{fancy}
\fancyhead{} % clear all header fields
\fancyhead[EC]{Yao-Zhong Zhang}
\fancyhead[EL,OR]{\thepage}
\fancyhead[OC]{Super Coherent States, Boson-Fermion Realizations and 
Representations...}
\fancyfoot{} % clear all footer fields
\renewcommand\headrulewidth{0.5pt}
\addtolength{\headheight}{2pt} % make space for the rule

%\begin{document}
%
%%\begin{titlepage}
%
%\begin{flushright}
%\end{flushright}
%
%%\baselineskip =16pt
%
%\vskip 1cm
%
%\begin{center}
%%\title
%{\Large\bf Super Coherent States, Boson-Fermion Realizations and 
%Representations of Superalgebras} \footnote{Invited contribution for the
%volume ``Progress in Field Theory Research" by Nova Science Publishers
%Inc., New York, 2004.}
%
%\vspace{1cm}
%
%%\author{
%{\large Yao-Zhong Zhang}
%\vskip.1in
%{\em Department of Mathematics, 
%University of Queensland, Brisbane, Qld 4072, Australia}
%
%\end{center}
%
%\date{}

%\maketitle

%\vspace{2cm}

\begin{flushleft}
\vspace{-3in}

\textit{\bfseries{Chapter 9}}
%\textit{\bfseries{Chapter 19}}

\end{flushleft}

\vspace{2.2in}

\begin{abstract}

Super coherent states are useful in the explicit construction of
representations of superalgebras and quantum superalgebras. In this 
contribution, we describe how they are used to construct (quantum)
boson-fermion realizations and representations of (quantum) superalgebras.
We work through a few examples: $osp(1|2)$ and its quantum
version $U_t[osp(1|2)]$, $osp(2|2)$ in the non-standard and
standard bases and $gl(2|2)$ in the non-standard basis. We obtain free
boson-fermion realizations of these superalgebras. Applying the boson-fermion
realizations, we explicitly construct their finite-dimensional representations.
Our results are expected to be useful in the study of current
superalgebras and their corresponding conformal field theories.

\end{abstract}

%\vspace{1cm}

%%%%%PACS: 11.25Hf; 11.30.Rd; 03.65Fd; 02.20.Hj.

%\vspace{0.5cm}

%\end{titlepage}

\setcounter{section}{0}
\setcounter{equation}{0}
\section{Introduction}

Supersymmetry concepts were used to provide a unified description of
mixed quantum mechanical systems of bosons and fermions. The
generators of supersymmetry transformations form a superalgebra. 
Superalgebras and their corresponding conformal field theories
(CFTs) have recently attracted considerable attention in mathematical physics
community, because of their applications
in areas such as topological field theory \cite{Roz92,Isi94},
logarithmic CFTs (see e.g. the review article \cite{Flohr01} and
references therein) and the supersymmetric method \cite{Efe83}
to disordered systems \cite{Ber95,Mud96,Maa97,Bas00,Gur00,Lud00,Bha01}.

Unlike ordinary bosonic algebras, there are two types of representations
for most superalgebras. They are the so-called typical and atypical
representations. The typical representations are irreducible and are
similar to the usual representations appeared in ordinary bosonic
algebras. The atypical representations have no counterpart in the
bosonic algebra representations. They can be irreducible or not
fully reducible (i.e. reducible or indecomposable). 
This makes the study of representations of superalgebras very difficult.

In most physical applications, one needs the explicit construction of
finite-dimensional representations of a superalgebra. This is
particularly the case in superalgebra CFTs. To construct primary fields
of such CFTs in terms of free fields, one has to construct the
finite-dimensional representations of the superalgebras explicitly.
The explicit construction of primary fields is essential in
the investigation of disordered systems by the supersymmetric method.

As is well-known, coherent states played an important role in a wide variety
of quantum mechanical problems and, in particular, in the explicit 
construction of representations of symmetry algebras. 
Super coherent states associated with superalgebras were introduced in 
\cite{Bal88,Bal89}. So it is natural to investigate representations of
superalgebras by means of super coherent states.

In the present contribution, we describe how the super
coherent state method can be used to explicitly construct finite-dimensional
representations of superalgebras. In this approach, the first step is to
obtain free boson-fermion realization of a superalgebra. Then representations
are constructed out of particle states in a super-Fock space.
We work through a few examples: the rank 1 superalgebra $osp(1|2)$ and
its quantum version $U_t[osp(1|2)]$, the rank 2 superalgebra $osp(2|2)$
in the non-standard and standard bases and the rank 3 non-semisimple 
superalgebra $gl(2|2)$ in the non-standard basis.

\setcounter{section}{1}
\setcounter{equation}{0}

\section{Boson-Fermion Realizations and Representations of $osp(1|2)$
and $U_t[osp(1|2)]$}

In this section we work through two rank-one superalgebras, $osp(1|2)$ 
and its quantized version $U_t[osp(1|2)]$,
to illustrate how to use the vector coherent state approach to
construct representations of superalgebras. Generalization to higher
rank cases will be disscussed in the following sections.

The superalgebra $osp(1|2)$ is generated by the odd simple elements $e,f$ and
even non-simple elements $E,F$. These elements obey the following
(anti-)commutation relations:
\beqa
&&\{e,f\}=H,~~~[H,e]=e,~~~[H,f]=-f,~~~\{e,e\}=2E,~~~\{f,f\}=-2F,\n
&&[E,F]=H,~~~[H,E]=2E,~~~[H,F]=-2F,~~~[e,F]=f,~~~[E,f]=-e.
\eeqa
The Casimir operator of this superalgebra is
\beq
C_2=\frac{1}{2}H(H+1)+fe+2FE.
\eeq

Before we proceed to the vector coherent state method, we write down
the Zassenhaus formula for two operators $X$ and $Y$, which will be
used frequently in the following. The general Zassenhaus formula reads
\beq
\exp(X+Y)=\exp(X)\exp(Y)\prod_{i=2}^\infty\,\exp(C_i),
\eeq
where $C_i$ is a homogeneous polynomial of order $i$ in $X$ and $Y$
and all $C_i$'s contain the commutator $[Y,X]$. The $C_i$'s up to sixth order
have been determined in \cite{Que03}. For our purpose, we only need
$C_i$'s up to fourth order. They are given by
\beqa
C_2&=&\frac{1}{2}[Y,X],~~~~C_3=\frac{1}{3}[[Y,X],Y]+\frac{1}{6}[[Y,X],X],\n
C_4&=&\frac{1}{8}\lt([[[Y,X],Y],Y]+[[[Y,X],X],Y]\rt)+\frac{1}{24}
    [[[Y,X],X],X].\label{Zassenhaus}
\eeqa

Let $|hw>$ be the highest weight state of $osp(1|2)$
with highest weight $j$ being a non-negative integer or half-integer, i.e.
\beq
H|hw>=2j|hw>,~~~~e|hw>=E|hw>=0.
\eeq
Define vector coherent states by $e^{f\a+aF}|hw>$, where $\a ~(\a^\dg)$ is 
a fermion destroying (creating) operator with number operator $N_\a$ and
$a~(a^\dg)$ is boson annhilation (creation) operator with number
operator $N_a$. They obey relations
\beqa
&&\{\a,\a^\dg\}=1,~~~[N_\a,\a]=-\a,~~~[N_\a,\a^\dg]=\a^\dg,~~~
   \a^2=\lt(\a^\dg\rt)^2=0,\n
&&[a,a^\dg]=1,~~~[N_a,a]=-a,~~~[N_a,a^\dg]=a^\dg.
\eeqa
Then state vectors $|\psi>$  are mapped into functions
\beq
\psi_j=<hw|e^{\a^\dg e+a^\dg E}|\psi>|0>=<hw|e^{\a^\dg e}e^{a^\dg E}|\psi>|0>.
\eeq
Here the vacuum vector $|0>$ is defined by $\a |0>=a|0>=0$.

Operators $A$ are mapped as follows:
\beq
A|\psi>\rtarr \G(A)\psi_j=<hw|e^{\a^\dg e}e^{a^\dg E}A|\psi>|0>.
\eeq
Taking $H,e,f,\cdots$ in turn we find
\beqa
\G(H)&=&2j-N_\a-2N_a,~~~\G(e)=\a+\a^\dg a,~~~\G(f)=2j\a^\dg-a^\dg\a
    -\a^\dg N_a,\n
\G(E)&=&a,~~~~\G(F)=a^\dg\lt(2j-N_a-N_\a\rt).
\eeqa
This gives rise to a boson-fermion realization of $osp(1|2)$. In this
realization, the Casimir takes a constant value, $C_2=2j(j+1/2)$.

The representations of $osp(1|2)$ are constructed as follows by using the
boson-fermion realization in the super-Fock space. 
First note that there two independent
combinations of the creation operators on the vacuum vector $|0>$:
$\lt(a^\dg\rt)^{j-m}|0>,~~j-m\in{\bf Z}_+$ and
$\a^\dg\lt(a^\dg\rt)^{j-m-1/2}|0>,~~j-m-1/2\in{\bf Z}_+$. This implies
that each representation of $osp(1|2)$ decomposes into at most two
representations of its even subalgebra $su(2)$. It is easy to check
these two states are already representations of $su(2)$ with higest
weights $j$ and $j-1/2$, respectively. This justifies the use of
notation $|j,m>$ and $|j-1/2,m>$ for the two $su(2)$ multiplets,
respectively:
\beqa
|j,m>&=&\lt(a^\dg\rt)^{j-m}|0>,~~~~m=j,j-1,\cdots,-j,\n
|j-\frac{1}{2},m>&=&\a^\dg\lt(a^\dg\rt)^{j-m-\frac{1}{2}}|0>,~~~~
   m=j-\frac{1}{2},j-\frac{3}{2},\cdots,-(j-\frac{1}{2}).
   \label{osp12-multiplets}
\eeqa
These two $su(2)$ multiplets span a spin $j$ representation of
$osp(1|2)$, to be denoted by $\pi_j$. The actions of the odd generators
on the multiplets (\ref{osp12-multiplets}) can be easily worked out,
which are not written down here. We denote by
$\sigma_j$ and $\sigma_{j-1/2}$ the two $su(2)$ representations in 
(\ref{osp12-multiplets}), respectively.
Then the $osp(1|2)\downarrow su(2)$ branching rule is
\beq
\pi_j=\sigma_j\oplus \sigma_{j-1/2}.
\eeq
The dimension of $\pi_j$ is $(2j+1)+2j=4j+1$. All representations
of $osp(1|2)$ are typical.

Now we generalize the method to quantum superalgebra $U_t[osp(1|2)]$.
The relations satisfied by the generators of $U_t[osp(1|2)]$ are given
by
\beqa
&&\{e,f\}=\frac{t^H-t^{-H}}{t-t^{-1}},~~~[H,e]=e,~~~[H,f]=-f,\n
&&\{e,e\}=2E,~~~\{f,f\}=-2F,~~~[H,E]=2E,~~~[H,F]=-2F,\n
&&[E,F]=ef\frac{t^{H-1/2}+t^{-H+1/2}}{t^{1/2}+t^{-1/2}}
   +fe\frac{t^{H+1/2}+t^{-H-1/2}}{t^{1/2}+t^{-1/2}},\n
&&[e,F]=f\frac{t^{H-1/2}+t^{-H+1/2}}{t^{1/2}+t^{-1/2}},~~~
  [E,f]=-e\frac{t^{H+1/2}+t^{-H-1/2}}{t^{1/2}+t^{-1/2}}.
\eeqa
Note that the right hand side of the 3rd equation is not equal to
$\frac{t^H-t^{-H}}{t-t^{-1}}$ but it degenerates to $H$ when $t=1$.
It follows that the $su(2)$ subalgebra of $osp(1|2)$
is not deformed into the usual quantum algebra $U_t[su(2)]$. Thus
the usual $U_t[su(2)]$ is not a subalgebra of $U_t[osp(1|2)]$.
This is the case for most superalgebras.

Define $t$-fermion operators $\a,\a^\dg$ and $t$-boson operators
$a,a^\dg$ with number operators $N_\a$ and $N_a$, respectively. They
satisfy the following relations:
\beqa
&&\a\a^\dg+t\a^\dg\a=t^{N_\a},~~~[N_\a,\a]=-\a,~~~[N_\a,\a^\dg]=\a^\dg,\n
&&\a^2=0=\lt(\a^\dg\rt)^2,~~~~\a^\dg\a=[N_\a]_t,~~~\a\a^\dg=[1-N_\a]_t,\n
&&aa^\dg-ta^\dg a=t^{-N_a},~~~[N_a,a]=-a,~~~[N_a,a^\dg]=a^\dg,\n
&&a^\dg a=[N_a]_t,~~~aa^\dg=[N_a+1]_t,
\eeqa
where
\beq
[x]_t\equiv\frac{t^x-t^{-x}} {t-t^{-1}}.
\eeq
Then state vectors are mapped into functions:
\beq
\psi_{t,j}=<hw|\exp(\a^\dg e)\exp_t(a^\dg E)|\psi>|0>,
\eeq
where $\exp_t(x)\equiv\sum_{n=0}^\infty\,\frac{x^n}{[n]_t!}$ is the
$t$-exponential, and operators $A$ are mapped as
\beq
A|\psi>\rtarr \G(A)\psi_{t,j}=<hw|\exp(\a^\dg e)\exp_t(a^\dg E)A|\psi>|0>.
   \label{mapping1}
\eeq
By means of the following $t$-identities,
\beqa
&&a\lt(a^\dg\rt)^n=t^n\lt(a^\dg\rt)^n
   a+[n]_t\lt(a^\dg\rt)^{n-1}t^{-N_a},\n
&&E^nf=fE^n-[n]_t\,e
   E^{n-1}\frac{t^{H+n-1/2}+t^{-H-n+1/2}}{t^{1/2}+t^{-1/2}},
\eeqa
we find from (\ref{mapping1}) a $t$-boson-fermion realization of 
$U_t[osp(1|2)]$:
\beqa
\G(H)&=&2j-N_\a-2N_a,~~~\G(e)=\a+\a^\dg a,\n
\G(f)&=&[2j]_t\a^\dg-\{4j-2N_a+1\}_{t}a^\dg\a
    -\{4j-2N_a-1\}_{t}\a^\dg [N_a]_t,\n
\G(E)&=&\{2N_\a-1\}_{t}a,\n
\G(F)&=&\{4j-2N_a+1\}_{t}\,a^\dg\lt([2j]_t\{2N_\a-1\}_{t}
   -[N_a]_t\{4j-2N_a-1\}_{t}[1-N_\a]_t\rt.\n
& &\lt.  -[N_a+1]_t\{4j-2N_a-3\}_{t}[N_\a]_t\rt),\label{t-realization}
\eeqa
where
\beq
\{x\}_{t}\equiv\frac{t^{\frac{x}{2}}+t^{-\frac{x}{2}}}{t^{\frac{1}{2}}
   +t^{-\frac{1}{2}}}.
\eeq
It can be checked that (\ref{t-realization}) indeed satisfies the
defining relations of $U_t[osp(1|2)]$.

Let us remark that boson-fermion models for $osp(1|2)$ and
$U_t[osp(1|2)]$ have been
obtained in \cite{Bra93} by using two bosons and one fermion. Our
realizations given above are different as they only involve one boson 
and one fermion.

Representations of $U_t[osp(1|2)]$ may be constructed by means of the
$t-$boson-fermion realization (\ref{t-realization}). 
The two multiplets for the unusual
deformation of the even subalgebra $su(2)$ of $osp(1|2)$ 
have the same form as (\ref{osp12-multiplets})
but now $a^\dg$ ($\a^\dg$) is the $t-$ boson (fermion) operator. This is
easily seen by the actions of $\G(E)$ and $\G(F)$ on the two
multiplets:
\beqa
\G(E)|j,m>&=&[j-m]_t\,|j,m+1>,\n
\G(F)|j,m>&=&[j+m]_t\{2(j+m-\frac{1}{2})\}_t\{2(j-m+\frac{1}{2})\}_t\,
   |j,m-1>,\n
\G(E)|j-\frac{1}{2},m>&=&[j-m-\frac{1}{2}]_t\,|j-\frac{1}{2},m+1>,\n
\G(F)|j-\frac{1}{2},m>&=&[j+m-\frac{1}{2}]_t\{2(j+m-\frac{1}{2})\}_t
   \{2(j-m+1)\}_t\, |j-\frac{1}{2},m-1>.\n
\eeqa
The actions of the odd generators on the two multiplets are given by
\beqa
\G(e)|j,m>&=&[j-m]_t\,|j-\frac{1}{2},m+\frac{1}{2}>,\n
\G(f)|j,m>&=&[j+m]_t\{2(j-m+\frac{1}{2})\}_t\,\,|j-\frac{1}{2},m-\frac{1}{2}>,\n
\G(e)|j-\frac{1}{2},m>&=&|j,m+\frac{1}{2}>,\n
\G(f)|j-\frac{1}{2},m>&=&-\{2(j+m)\}_t\,|j,m-\frac{1}{2}>.
\eeqa

\section{Boson-Fermion Realizations of $osp(2|2)$}

Superalgebra $osp(2|2)$ can be written as $osp(2|2)=
osp(2|2)^{\rm even}\oplus osp(2|2)^{\rm odd}$, with
\beq
osp(2|2)^{\rm even}=u(1)\oplus su(2)=\{H'\}\oplus\{H, E, F\},~~~~
osp(2|2)^{\rm odd}=\{e,f,\bar{e},\bar{f}\},
\eeq
where $e,\;f,\;\bar{e},\;\bar{f}$ are the generators corresponding to the
fermionic roots, and $E,\;F$ are those to the bosonic roots.
It is well-known that unlike a purely bosonic algebra 
a superalgebra admits different Weyl inequivalent
choices of simple root systems, which correspond to inequivalent Dynkin 
diagrams. In the case of $osp(2|2)$, one has two choices of simple root
systems which are unrelated by Weyl transformations: a system of fermionic and
bosonic simple roots (i.e. the so-called standard basis), or a purely
fermionic system of simple roots (that is the so-called non-standard
basis). So it is useful to obtain results in the two different bases
for different physical applications (see e.g. the discussions in
\cite{Bow96,Ras98,Ding03}). 

\subsection{Realization in the Non-standard Basis}

In the non-standard basis, simple roots of $osp(2|2)$ are all fermionic.
Let $e,\;f,\;\bar{e},\;\bar{f}$ be the generators corresponding to such
fermionic simple roots, and let $E,\;F$ be the non-simple generators.
They obey the (anti-) commutation relations:
\beqa
 &&\{e,f\}=-\frac{1}{2}(H-H^{\prime}), ~~~~[H,e]=e,~~~~[H,f]=-f, \n
 &&[H^{\prime},e]=e,~~~~[H^{\prime},f]=-f,\n
 &&\{\bar{e},\bar{f}\}=-\frac{1}{2}(H+H^{\prime}),~~~~
    [H,\bar{e}]=\bar{e},~~~~[H,\bar{f}]=-\bar{f},\n
 &&[H^{\prime},\bar{e}]=-\bar{e},~~~~
    [H^{\prime},\bar{f}]=\bar{f},\n
 &&\{e,\bar{e}\}=E,~~~~\{\bar{f},f\}=-F, \n
 &&[E,F]=H, ~~~~[H,E]=2E,~~~~[H,F]=-2F, \n
 &&[E,f]=\bar{e},~~~~[F,e]=\bar{f},~~~~ 
   [E,\bar{f}]=e,~~~~[F,\bar{e}]=f.  \label{cr-nst}
\eeqa

\no All other (anti-)commutators are zero, and the quadratic Casimir is 
\beq
C_2=\frac{1}{2}\left(H^2- H^{\prime~2}\right) 
-2fe -2 {\bar f}{\bar e}+2FE.
\eeq

Let $|hw>$ be the highest weight state of $osp(2|2)$ in the non-standard
basis and $(J,q)$ be the corresponding highest weight. Namely,
\beq
H|hw>=2J|hw>,~~~~H'|hw>=2q|hw>,~~~~E|hw>=e|hw>=\bar{e}|hw>=0.
\eeq
Define the vector coherent states, $e^{Fa+f\a_1+\a_2\bar{f}}|hw>$.
Then state vectors $|\psi>$ are mapped into functions
\beqa
\psi_{J,q}&=&<hw|\exp\lt(\a^\dg_1 e+\a^\dg_2 \bar{e}+a^\dg E\rt)|\psi>|0>\n
&=&<hw|e^{(a^\dg+\frac{1}{2}\a_1^\dg\a_2^\dg)E}e^{\a_1^\dg e}
   e^{\a_2^\dg\bar{e}}|\psi>|0>.
\eeqa
Here $a, a^\dg$ are bosonic operators with number operator $N_a$, and
$\a_1~(\a_1^\dg), \a_2~(\a_2^\dg)$ are fermionic operators with number
operators $N_{\a_1}, N_{\a_2}$, respectively. These operators satisfy
relations:
\beqa
&&[a, a^\dg]=1,~~~~[N_a, a^\dg]=a^\dg,~~~~[N_a, a]=-a,\n
&&\{\a_1,\a_1^\dg\}=1,~~~~[N_{\a_1}, \a_1^\dg]=\a_1^\dg,~~~~[N_{\a_1},
   \a_1]=-\a_1,\n
&&\{\a_2,\a_2^\dg\}=1,~~~~[N_{\a_2}, \a_2^\dg]=\a_2^\dg,~~~~[N_{\a_2},
   \a_2]=-\a_2,
\eeqa
all other (anti-)commutators are zero. Moreover,
$a|0>=\a_1|0>=\a_2|0>=0$.

Operators $A$ in the non-standard basis are mapped as before:
\beq
A|\psi>\rtarr \G(A)\psi_{J,q}=<hw|e^{(a^\dg+\frac{1}{2}\a_1^\dg\a_2^\dg)E}
e^{\a_1^\dg e}e^{\a_2^\dg\bar{e}}A|\psi>|0>.
\eeq
Taking $A=H, H', e, \cdots$ in turn we find
\beqa
\G(H)&=&2J-2N_a-N_{\a_1}-N_{\a_2},~~~~\G(H')=2q-N_{\a_1}+N_{\a_2},\n
\G(e)&=&\a_1+\frac{1}{2}\a_2^\dg a,~~~~\G(f)=-(J-q)\a_1^\dg+a^\dg \a_2
   +\frac{1}{2}\a_1^\dg\lt(N_a+N_{\a_2}\rt),\n
\G(\bar{e})&=&\a_2+\frac{1}{2}\a_1^\dg a,~~~~\G(\bar{f})=-(J+q)\a_2^\dg
   +a^\dg \a_1+\frac{1}{2}\a_2^\dg\lt(N_a+N_{\a_1}\rt),\n
\G(E)&=&a,~~~~\G(F)=2Ja^\dg+q\a_1^\dg\a_2^\dg-a^\dg\lt(N_a+N_{\a_1}
   +N_{\a_2}\rt).\label{realization-nst}
\eeqa
This gives a free boson-fermion realization of $osp(2|2)$ in the
non-standard basis.  In this realization, the Casimir takes a constant
value, i.e. $C_2=2(J^2-q^2)$.

\subsection{Realization in the Standard Basis}
 
Let $E\;(F)$ and $e\;(f)$ be the generators 
corresponding to the even and odd simple roots of $osp(2|2)$
in the standard (distinguished) basis, 
respectively. Let $\bar{e},\; \bar{f}$ be the odd non-simple generators. 
They satisfy the following (anti-)commutation relations: 
\beqa
 &&[E,F]=H, ~~~~[H,E]=2E,~~~~[H,F]=-2F, \n
 &&\{e,f\}=-\frac{1}{2}(H-H^{\prime}), ~~~~[H,e]=-e,~~~~[H,f]=f, \n
 &&[H^{\prime},e]=-e,~~~~[H^{\prime},f]=f,\n
 &&[E,e]=\bar{e},~~~~[F,f]=\bar{f}, \n
 &&\{\bar{e},\bar{f}\}=-\frac{1}{2}(H+H^{\prime}),~~~~
   [H,\bar{e}]=\bar{e},~~~~[H,\bar{f}]=-\bar{f},\n
 &&[H^{\prime},\bar{e}]=-\bar{e},~~~~[H^{\prime},\bar{f}]=\bar{f},\n
 &&\{e,\bar{f}\}=-F,~~~~\{\bar{e},f\}=E,~~~~ 
   [E,\bar{f}]=f,~~~~[F,\bar{e}]=e.
\eeqa

\no All other (anti-)commutators are zero. The quadratic Casimir 
is given by
\beq
C_2=\frac{1}{2}\left(H(H+2)- H^{\prime}(H^{\prime}+2)\right) 
+ 2fe -2 {\bar f}{\bar e}+2FE.
\eeq

Let $|hw>$ be the highest weight state of highest weight $(p,q)$
of $osp(2|2)$ in the standard basis:
\beq
H|hw>=2p|hw>,~~~~H'|hw>=2q|hw>,~~~~E|hw>=e|hw>=\bar{e}|hw>=0.
\eeq
Then similar to the non-standard basis case,
state vectors $|\psi>$ in the standard basis are mapped into functions
\beqa
\psi_{p,q}&=&<hw|\exp\lt(a^\dg E+\a^\dg_1 e+\a^\dg_2 \bar{e}\rt)|\psi>|0>\n
&=&<hw|e^{a^\dg E}e^{\a_1^\dg e}
   e^{(\a_2^\dg-\frac{1}{2}\a_1^\dg a^\dg)\bar{e}}|\psi>|0>,
\eeqa
and operators $A$ are mapped as before:
\beq
A|\psi>\rtarr \G(A)\psi_{J,q}=<hw|e^{a^\dg E}e^{\a_1^\dg e}
e^{(\a_2^\dg-\frac{1}{2}\a_1^\dg a^\dg)\bar{e}}A|\psi>|0>.
\eeq
After some algebraic manipulations, we find
\beqa
\G(H)&=&2p-2N_a+N_{\a_1}-N_{\a_2},~~~~\G(H')=2q+N_{\a_1}+N_{\a_2},\n
\G(E)&=&a-\frac{1}{2}\a_1^\dg\a_2,\n
\G(F)&=&2pa^\dg-\a_2^\dg\a_1-a^\dg \lt(N_a-\frac{1}{2}N_{\a_1}+\frac{1}{2}
    N_{\a_2}\rt)-\frac{1}{4}(a^\dg)^2\a_1^\dg\a_2,\n
\G(e)&=&\a_1+\frac{1}{2}a^\dg \a_2,~~~~\G(f)=-(p-q)\a_1^\dg+\a_2^\dg a
   +\frac{1}{2}\a_1^\dg\lt(N_a+N_{\a_2}\rt),\n
\G(\bar{e})&=&\a_2,~~~~\G(\bar{f})=-(p+q)\a_2^\dg
   -\frac{1}{2}(3p-q)a^\dg \a_1^\dg+\a_2^\dg\lt(N_a-N_{\a_1}\rt)
   +\frac{1}{2}a^\dg\a_1^\dg N_a.\n\label{realization-st}
\eeqa
This is the free boson-fermion realization of $osp(2|2)$ in the
standard basis.  In this realization, $C_2=2[p(p+1)-q(q+1)]$.

\section{Construction of Representations of $osp(2|2)$}

We now use the above free boson-fermion realizations to construct 
finite-dimensional representations of
$osp(2|2)$ in both the non-standard and standard bases. 
As we will see, all finite-dimensional typical and atypical
representations of $osp(2|2)$ can be constructed in an unified manner.

\subsection{Representations in the Non-standard Basis}

Representations of $osp(2|2)$ in the non-standard basis were
investigated in \cite{Sch77,Mar80}. Here we reproduce those results by means of
the free boson-fermion realization obtained in the previous section.
Our method is different from, and in our view simpler than, the
methods used in \cite{Sch77,Mar80}.
To begin with, we note that representations of $osp(2|2)$ in the
non-standard basis are labelled by $(J,q)$, where $J$ is a non-negative
integer or half-integer and $q$ is an arbitrary complex number.
There are four independent combinations of
creation operators acting on the vacuum vector $|0>$:
\beqa
&&\lt(a^\dg\rt)^{J-m}|0>,~~~J-m\in {\bf Z}_+,\n
&&\a_1^\dg\lt(a^\dg\rt)^{J-m-1/2}|0>,~~~J-\frac{1}{2}-m\in {\bf Z}_+,\n
&&\a_2^\dg\lt(a^\dg\rt)^{J-m-1/2}|0>,~~~J-\frac{1}{2}-m\in {\bf Z}_+,\n
&&\a_1^\dg\a_2^\dg\lt(a^\dg\rt)^{J-m-1}|0>,~~~J-1-m\in {\bf Z}_+.
\eeqa
Thus each $osp(2|2)$ representation decomposes into at most four
representations of the even subalgebra $su(2)\oplus u(1)$. Let us
construct representations for $su(2)\oplus u(1)$ out of the above
particle states in the super-Fock space. It is easy to check that 
the last three states are already representations of
$su(2)\oplus u(1)$ with highest weights $(J-\frac{1}{2}, q-\frac{1}{2}),
~(J-\frac{1}{2}, q+\frac{1}{2})$ and $(J-1, q)$, respectively. This
justifies the use of the notations, $|J-\frac{1}{2},m;q-\frac{1}{2}>,~
|J-\frac{1}{2},m;q+\frac{1}{2}>$ and $|J-1,m;q>$ for these three multiplets,
respectively. To turn the first state into a representation of
$su(2)\oplus u(1)$, we make the following ansatz in view
of the free boson-fermion expressions of generators $\G(H'), \G(H), \G(E), 
\G(F)$,
\beq
\eta_{J,q}^m=r_{J,q}^m\lt(a^\dg\rt)^{J-m}|0>+\bar{r}_{J,q}^m
   \a_1^\dg\a_2^\dg\lt(a^\dg\rt)^{J-m-1} |0>,
\eeq
where $r_{J,q}^m, \bar{r}_{J,q}^m$ are functions of $J,q,m$ to
be determined. It is easily shown that
$\G(H)\eta_{J,q}^m=2m\eta_{J,q}^m$ and $\G(H')\eta_{J,q}^m=2q\eta_{J,q}^m$.
So this state has highest weight $(J,q)$ and thus will be written as
$|J,q;m>$ in the following. Now
\beq
\G(E)|J,m;q>=(J-m)r_{J,q}^m\lt(a^\dg\rt)^{J-(m+1)}|0>+(J-m-1)\bar{r}_{J,q}^m
   \a_1^\dg\a_2^\dg\lt(a^\dg\rt)^{J-(m+1)-1} |0>.
\eeq
The r.h.s. must equal to $(J-m)|J,m+1;q>$ for the representation to be
finite-dimensional. It follows that
\beqa
r_{J,q}^m=r_{J,q}^{m+1} &\Longrightarrow& r_{J,q}^m=c_{J,q},\n
(J-m)\bar{r}_{J,q}^{m+1}=(J-m-1)\bar{r}_{J,q}^m &\Longrightarrow&
    \bar{r}_{J,q}^m=(J-m)\bar{c}_{J,q},
\eeqa
where $c_{J,q}, \bar{c}_{J,q}$ are functions of $J,q$ only.
Finally the action of $\G(F)$ on $|J,q;m>$ gives
\beqa
\G(F)|J,m;q>&=&c_{J,q}(J+m)\lt(a^\dg\rt)^{J-(m-1)} |0>\n
&+& [c_{J,q} q+\bar{c}_{J,q}(J-m)(J+m-1)]\a_1^\dg\a_2^\dg
   \lt(a^\dg\rt)^{J-(m-1)-1} |0>.\n
\eeqa
This becomes a finite-dimensional representation of the even subalgebra
if one requires that the r.h.s equal to $(J+m)|J,m-1;q>$. This requirement 
is satisfied if $c_{J,q}=2J$ and $\bar{c}_{J,q}=q$. So 
representations of  $osp(2|2)$ are spanned by the $su(2)\oplus u(1)$
multiplets:
\beqa
&&|J,m;q>=\lt[2J\lt(a^\dg\rt)^{J-m}+q(J-m)\a_1^\dg\a_2^\dg\lt(a^\dg\rt)^{J-m-1}
   \rt]|0>,~~~m=J,J-1,\cdots, -J,\n
&&|J-\frac{1}{2},m;q-\frac{1}{2}>=\a_1^\dg\lt(a^\dg\rt)^{J-1/2-m}|0>,~~~
   m=J-\frac{1}{2}, J-\frac{3}{2},\cdots,-(J-\frac{1}{2}),\n
&&|J-\frac{1}{2},m;q+\frac{1}{2}>=\a_2^\dg\lt(a^\dg\rt)^{J-1/2-m}|0>,~~~
   m=J-\frac{1}{2}, J-\frac{3}{2},\cdots,-(J-\frac{1}{2}),\n
&&|J-1,m;q>=\a_1^\dg\a_2^\dg\lt(a^\dg\rt)^{J-1-m}|0>,~~~
   m=J-1, J-2,\cdots,-(J-1).\label{osp22-nst-mts}
\eeqa
Here $J\geq 1/2$ in the first three expressions and $J\geq 1$ in the last
one.  Note that the trivial 1-dimensional representation of $osp(2|2)$ 
(corresponding to $J=0=q$) is given by the vacuum $|0>$.
It follows that in the non-standard basis, $osp(2|2)\downarrow
su(2)\oplus u(1)$ branching rule is 
\beq
\pi_{(J,q)}=\sigma_{(J,q)}\oplus\sigma_{(J-1/2,q-1/2)}\oplus
   \sigma_{(J-1/2,q+1/2)}\oplus\sigma_{(J-1,q)}
\eeq
for $q\neq \mp J$. Here $\pi_{(J,q)}$ stands for a representation of
$osp(1|2)$ (in the non-standard basis) labelled by $(J,q)$, and
$\sigma_{(J,q)}$ etc stand for the four $su(2)\oplus u(1)$ multiplets
in (\ref{osp22-nst-mts}).

By means of (\ref{realization-nst}) and (\ref{osp22-nst-mts}) it
can be shown that the actions of the odd generators on 
these $su(2)\oplus u(1)$ multiplets are given by
\beqa
\G(e)|J,m;q>&=&(J-m)(q+J)|J-\frac{1}{2},m+\frac{1}{2};q+\frac{1}{2}>,\n
\G(f)|J,m;q>&=&(J+m)(q-J)|J-\frac{1}{2},m-\frac{1}{2};q-\frac{1}{2}>,\n
\G(\bar{e})|J,m;q>&=&-(J-m)(q-J)|J-\frac{1}{2},m+\frac{1}{2};q-\frac{1}{2}>,\n
\G(\bar{f})|J,m;q>&=&-(J+m)(q+J)|J-\frac{1}{2},m-\frac{1}{2};q+\frac{1}{2}>,
\eeqa
\beqa
\G(\bar{e})|J-\frac{1}{2},m;q-\frac{1}{2}>&=&0,\n
\G(\bar{f})|J-\frac{1}{2},m;q-\frac{1}{2}>&=&\frac{1}{2J}|J,m-\frac{1}{2};q>
  -\frac{J-1/2+m}{2J}(q+J)|J-1,m-\frac{1}{2}; q>,\n
\G(e)|J-\frac{1}{2},m;q-\frac{1}{2}>&=&\frac{1}{2J}|J,m+\frac{1}{2};q>
  -\frac{J-1/2-m}{2J}(q+J)|J-1,m+\frac{1}{2}; q>,\n
\G(f)|J-\frac{1}{2},m;q-\frac{1}{2}>&=&0,
\eeqa 
\beqa
\G(e)|J-\frac{1}{2},m;q+\frac{1}{2}>&=&0,\n
\G(f)|J-\frac{1}{2},m;q+\frac{1}{2}>&=&\frac{1}{2J}|J,m-\frac{1}{2};q>
  +\frac{J-1/2+m}{2J}(q-J)|J-1,m-\frac{1}{2}; q>,\n
\G(\bar{e})|J-\frac{1}{2},m;q+\frac{1}{2}>&=&\frac{1}{2J}|J,m+\frac{1}{2};q>
  -\frac{J-1/2-m}{2J}(q-J)|J-1,m+\frac{1}{2}; q>,\n
\G(\bar{f})|J-\frac{1}{2},m;q+\frac{1}{2}>&=&0,
\eeqa 
and
\beqa
\G(e)|J-1,m;q>&=&|J-\frac{1}{2},m+\frac{1}{2};q+\frac{1}{2}>,\n
\G(f)|J-1,m;q>&=&-|J-\frac{1}{2},m-\frac{1}{2};q-\frac{1}{2}>,\n
\G(\bar{e})|J-1,m;q>&=&-|J-\frac{1}{2},m+\frac{1}{2};q-\frac{1}{2}>,\n
\G(\bar{f})|J-1,m;q>&=&|J-\frac{1}{2},m-\frac{1}{2};q+\frac{1}{2}>,
\eeqa
Note that the multiplet $|J,m;q>$ has dimension $2J+1$, both
$|J-\frac{1}{2}, m;q\mp\frac{1}{2}>$ have dimension $2J$ and the
dimension of $|J-1,m;q>$ is $2J-1$. So for $q\neq \mp J$,
they constitute irreducible
typical representation of dimension $8J$ of $osp(2|2)$.

When $q=\mp J$, the representations become atypical. We have different types
of atypical representations. The Casimirs for all such representations
vanish, and yet they are not the trivial one-dimensional representation. 
One type is obtained by applying the odd generators to the $su(2)\oplus
u(1)$ representation $|J,m;q>$. 
%droping either the multiplets
%$|J-\frac{1}{2},m;q-\frac{1}{2}>$ and $|J-1,m;q>$ or the multiplets
%$|J-\frac{1}{2},m;q+\frac{1}{2}>$ and $|J-1,m;q>$. 
For $q=J$, the multiplets
$|J-\frac{1}{2},m;q-\frac{1}{2}>$ and $|J-1,m;q>$ do not
appear, and only $|J,m;q>$ and $|J-\frac{1}{2},m;q+\frac{1}{2}>$ survive.
They form irreducible atypical representation of dimension $4J+1$. Similarly,
for $q=-J$, the multiplets $|J-\frac{1}{2},m;q+\frac{1}{2}>$ and $|J-1,m;q>$
do not appear, and only $|J,m;q>$ and $|J-\frac{1}{2},m;q-\frac{1}{2}>$ remain.
They also form irreducible atypical representation of dimension $4J+1$.
Other types of atypical representations  are not irreducible and are
obatined by applying odd generators to the $su(2)\oplus u(1)$
representations
$|J-\frac{1}{2},m;q\mp \frac{1}{2}>$. In both cases, the representation
contains all multiplets and thus has dimension $8J$. These
representations are not fully reducible.

\subsection{Representations in the Standard Basis}

Representations of $osp(2|2)$ in the standard basis are labelled by
$(p,q)$ with $p$ being a non-negative integer or half-integer and $q$ any
complex number.

In the standard basis,  the four independent combinations of
creation operators acting on~$|0>$ are
\beqa
&&\lt(a^\dg\rt)^{p-m}|0>,~~~p-m\in {\bf Z}_+,\n
&&\a_1^\dg\lt(a^\dg\rt)^{p-m-1/2}|0>,~~~p-\frac{1}{2}-m\in {\bf Z}_+,\n
&&\a_2^\dg\lt(a^\dg\rt)^{p-m-3/2}|0>,~~~p-\frac{3}{2}-m\in {\bf Z}_+,\n
&&\a_1^\dg\a_2^\dg\lt(a^\dg\rt)^{p-m-2}|0>,~~~p-2-m\in {\bf Z}_+.
    \label{com-st}
\eeqa
So again each $osp(2|2)$ representation decomposes into at most four
representations of the even subalgebra $su(2)\oplus u(1)$. To construct
the four multiplets, we note that the first and
the last states are already representations of $su(2)\oplus u(1)$ with highest
weight weights $(p,q)$ and $(p,q+1)$, respectively. We denote these two
multiplets by $|p,m;q>$ and $|p,m;q+1>$, respectively. We can
combine the second and the third states into two independent
multiplets of $su(2)\oplus u(1)$ with highest weights $(p-\frac{1}{2},
q+\frac{1}{2})$ and $(p+\frac{1}{2},q+\frac{1}{2})$, respectively.
This is seen as follows. Let
\beq
\chi_{p,q}^m=\frac{1}{2}c_{p,q}^m\a_1^\dg\lt(a^\dg\rt)^{p-m-1/2}|0>
   +\bar{c}_{p,q}^m\a_2^\dg\lt(a^\dg\rt)^{p-m-3/2}|0>,
\eeq
where $c_{p,q}^m$ and $\bar{c}_{p,q}^m$ are functions of $p,q,m$ to be
determined. Then,
\beqa
\G(E)\chi_{p,q}^m&=&\frac{1}{2}\lt((p-m-\frac{1}{2})c_{p,q}^m-\bar{c}_{p,q}^m
  \rt)\a_1^\dg\lt(a^\dg\rt)^{p-m-3/2}|0>\n
& &+(p-m-\frac{3}{2})\bar{c}_{p,q}^m\a_2^\dg\lt(a^\dg\rt)^{p-m-5/2}|0>.
\eeqa
To make the representation finite-dimensional, the r.h.s. of this
equation must equal to
$(p-m-\frac{x}{2})\chi_{p,q}^{m+1}$ for some integer $x$. 
On the other hand,
\beqa
\G(F)\chi_{p,q}^m&=&\frac{1}{2}\lt((p+m+1)c_{p,q}^m-\frac{1}{2}
  \bar{c}_{p,q}^m\rt)\a_1^\dg\lt(a^\dg\rt)^{p-m+1/2}|0>\n
& &\lt((p+m+1)\bar{c}_{p,q}^m-\frac{1}{2}c_{p,q}^m\rt)
  \a_2^\dg\lt(a^\dg\rt)^{p-m-1/2}|0>.
\eeqa
This must equal to $(p+m+\frac{y}{2})\chi_{p,q}^{m-1}$ for some integer
$y$ in order for the representation to be finite-dimensional. 
These two requirements are satisfied if \cite{Zha03} either
\beq
x=1,~~~y=3,~~~c_{p,q}^m=3p+m+\frac{5}{2},~~~\bar{c}_{p,q}^m=-(p-m-\frac{1}{2})
   \label{case1}
\eeq
or
\beq
x=3,~~~y=1,~~~~c_{p,q}^m=\bar{c}_{p,q}^m=1. \label{case2}
\eeq
Also it is easily seen that
\beq
\G(H)\chi_{p,q}^m=2(m+1)\chi_{p,q}^m,~~~~\G(H')\chi_{p,q}^m=2(q+
\frac{1}{2})\chi_{p,q}^m.
\eeq
It follows that $\chi_{p,q}^m$ has highest weight $(p+1/2,q+1/2)$
for the first case  (\ref{case1}) 
(where $m_{\rm max}=p-1/2$) and highest weight
$(p-1/2,q+1/2)$ for the second case (\ref{case2})
(where $m_{\rm max}=p-3/2$). This justifies
the use of notation, $|p+\frac{1}{2},m;q+\frac{1}{2}>$ and
$|p-\frac{1}{2},m;q+\frac{1}{2}>$, for these two multiplets, respectively.

Therefore, we have the following four $su(2)\oplus u(1)$  multiplets
which span finite-dimensional representations of $osp(2|2)$:
\beqa
|p,m;q>&=&\lt(a^\dg\rt)^{p-m}|0>,~~~m=p,p-1,\cdots, -p,~~~p\geq 0,\n
|p-\frac{1}{2},m;q+\frac{1}{2}>&=&\lt(\a_2^\dg+\frac{1}{2}\a_1^\dg
   a^\dg\rt) \lt(a^\dg\rt)^{p-3/2-m}|0>,\n
& &   m=p-\frac{3}{2}, p-\frac{5}{2},\cdots,-(p+\frac{1}{2}),~~~
   p\geq \frac{1}{2},\n
|p+\frac{1}{2},m;q+\frac{1}{2}>&=&\lt(p+m+\frac{3}{2}\rt)\a_1^\dg
   \lt(a^\dg\rt)^{p-1/2-m}|0>\n
& &   -\lt(p-m-\frac{1}{2}\rt)\lt(\a_2^\dg-
   \frac{1}{2}\a_1^\dg a^\dg\rt)\lt(a^\dg\rt)^{p-m-3/2}|0>,\n
& &  m=p-\frac{1}{2}, p-\frac{3}{2},\cdots,-(p+\frac{3}{2}),~~~p\geq 0,\n
|p,m;q+1>&=&\a_1^\dg\a_2^\dg\lt(a^\dg\rt)^{p-2-m}|0>,\n
& &   m=p-2, p-3,\cdots,-(p+2),~~~p\geq 0.\label{osp22-st-mts}
\eeqa
We remark that the trivial 1-dimensional representation (for which
$p=0=q$) is provided by $|0>$. The $osp(2|2)\downarrow su(2)
\oplus u(1)$ branching rule in the standard basis is given by
\beq
\pi_{(p,q)}=\sigma_{(p,q)}\oplus\sigma_{(p-1/2,q+1/2)}\oplus
   \sigma_{(p+1/2,q+1/2)}\oplus\sigma_{(p,q+1)}
\eeq
for $q\neq p, -p-1$. Here $\pi_{(p,q)}$ is a representation of
$osp(2|2)$ (in the standard basis) labelled by $(p,q)$, and
$\sigma_{(p,q)}$ etc are the four $su(2)\oplus u(1)$ multiplets in
(\ref{osp22-st-mts}).

Using (\ref{realization-st}) and (\ref{osp22-st-mts}), one can compute
the actions of the odd generators on these $su(2)\oplus u(1)$ multiplets.
The result is given by \cite{Zha03}
\beqa
\G(e)|p,m;q>&=&0,\n
\G(f)|p,m;q>&=&\frac{p-m}{2p+1}(q+p+1)
    |p-\frac{1}{2},m-\frac{1}{2};q+\frac{1}{2}>\n
& &    +\frac{1}{2p+1}(q-p)|p+\frac{1}{2},m-\frac{1}{2};q+\frac{1}{2}>,\n
\G(\bar{e})|p,m;q>&=&0,\n
\G(\bar{f})|p,m;q>&=&-\frac{p+m}{2p+1}(q+p+1)
    |p-\frac{1}{2},m-\frac{3}{2};q+\frac{1}{2}>,\n
& & +\frac{1}{2p+1}(q-p)|p+\frac{1}{2},m-\frac{3}{2};q+\frac{1}{2}>,
\eeqa
\beqa
\G(e)|p-\frac{1}{2},m;q+\frac{1}{2}>&=&|p,m+\frac{1}{2};q>,\n
\G(f)|p-\frac{1}{2},m;q+\frac{1}{2}>&=&(q-p)|p,m-\frac{1}{2}; q+1>,\n
\G(\bar{e})|p-\frac{1}{2},m;q+\frac{1}{2}>&=&|p,m+\frac{3}{2};q>,\n
\G(\bar{f})|p-\frac{1}{2},m;q+\frac{1}{2}>&=&
  (q-p)|p,m-\frac{3}{2}; q+1>,
\eeqa 
\beqa
\G(e)|p+\frac{1}{2},m;q+\frac{1}{2}>&=&(p+m+\frac{3}{2})|p,m+\frac{1}{2};q>,\n
\G(f)|p+\frac{1}{2},m;q+\frac{1}{2}>&=&-(p-m-\frac{1}{2})(q+p+1)
   |p,m-\frac{1}{2};q+1>,\n
\G(\bar{e})|p+\frac{1}{2},m;q+\frac{1}{2}>&=&-(p-m-\frac{1}{2})
   |p,m+\frac{3}{2};q>,\n
\G(\bar{f})|p+\frac{1}{2},m;q+\frac{1}{2}>&=&(p+m+\frac{3}{2})(q+p+1)
   |p,m-\frac{3}{2};q+1>,
\eeqa 
and
\beqa
\G(e)|p,m;q+1>&=&\frac{p+m+2}{2p+1}
   |p-\frac{1}{2},m+\frac{1}{2};q+\frac{1}{2}>\n
& &-\frac{1}{2p+1}|p+\frac{1}{2},m+\frac{1}{2};q+\frac{1}{2}>,\n
\G(f)|p,m;q+1>&=&0,\n
\G(\bar{e})|p,m;q+1>&=&-\frac{p-m-2}{2p+1}
   |p-\frac{1}{2},m+\frac{3}{2};q+\frac{1}{2}>\n
& &-\frac{1}{2p+1}|p+\frac{1}{2},m+\frac{3}{2};q+\frac{1}{2}>,\n
\G(\bar{f})|p,m;q+1>&=&0,
\eeqa
Note that both $|p,m;q>$ and $|p,m;q+1>$ have dimension $2p+1$, 
$|p-\frac{1}{2}, m;q+\frac{1}{2}>$ has dimension $2p$ and the
dimension of $|p+\frac{1}{2}, m;q+\frac{1}{2}>$ is $2p+2$. So for $q\neq p,
-p-1$, they constitute irreducible
typical representation of dimension $8p+4$ of $osp(2|2)$.

When $q=p,~-p-1$, the representations become atypical. The Casimirs for
such representations vanish. For $q=p$, one drops $|p+\frac{1}{2},
m;q+\frac{1}{2}>$ and $|p,m;q+1>$ and only keeps $|p,m;q>$ and 
$|p-\frac{1}{2}, m;q+\frac{1}{2}>$ survive. They form irreducible atypical
representation of $osp(2|2)$ of dimension $4p+1~ (p\geq 1/2)$.
For $q=-p-1$, $|p-\frac{1}{2}, m;q+\frac{1}{2}>$ and $|p,m;q+1>$
disappear and only $|p,m;q>$ and $|p+\frac{1}{2}, m;q+\frac{1}{2}>$
remain. They constitute irreducible atypical representation of dimension
$4p+3$. Other atypical representations  are not irreducible and obtained by
retaining all the $su(2)\oplus u(1)$ multiplets. Such representation has
dimension $8p+4$ and is not fully reducible.

\section{Boson-Fermion Realizations of $gl(2|2)$ and Its Subalgebras}

In this section, we obtain boson-fermion realizations of the
superalgebra $gl(2|2)$ and its subalgebras in the non-standard basis.

This superalgebra is non-semisimple and can be written as 
$gl(2|2)^{\rm even}\oplus gl(2|2)^{\rm odd}$, where
\beqa
gl(2|2)^{\rm even}&=&u(1)\oplus gl(2)\oplus gl(2)\n
&=&\{I\}\oplus\{\{E_{12},E_{21},H_1\},\{E_{34},E_{43},H_2\},N\},\n
gl(2|2)^{\rm odd}&=&\{E_{13},E_{31},E_{32},E_{23},E_{24},E_{42},
   E_{14},E_{41}\}.
\eeqa
In the non-standard basis, $E_{13},E_{32},E_{24}$
($E_{31},E_{23},E_{42}$) are simple raising (lowering) generators, 
$E_{12},E_{34},E_{14}$ ($E_{21},E_{43},E_{41}$) are non-simple
raising (lowering) generators and $I,H_1,H_2,N$ are elements of
the Cartan subalgebra. We have  
\beqa
H_1&=&E_{11}-E_{22},~~~~H_2=E_{33}-E_{44},\n
I&=&E_{11}+E_{22}+E_{33}+E_{44},\n
N&=&E_{11}+E_{22}-E_{33}-E_{44}+\beta I\label{cartan-gl22}
\eeqa
with $\beta$ being an arbitrary parameter. That $N$ is not uniquely
determined is a consequence of the fact that $gl(2|2)$ is
non-semisimple. The generators obey the following (anti-)commutation
relations:
\beq
[E_{ij},E_{kl}]=\delta_{jk}E_{il}-(-1)^{([i]+[j])([k]+[l])}\delta_{il}
   E_{kj},
\eeq
where $[E_{ij},E_{kl}]\equiv
E_{ij}E_{kl}-(-1)^{([i]+[j])([k]+[l])}E_{kl}E_{ij}$ is a commutator or
an anticommutator, $[1]=[2]=0,~
[3]=[4]=1$, and $E_{ii},~i=1,2,3,4$ are related to $I,H_1,H_2,N$
via (\ref{cartan-gl22}).

Let $|hw>$ be the highest weight state of $gl(2|2)$ defined by
\beqa
&&H_1|hw>=2J_1|hw>,~~~H_2|hw>=2J_2|hw>,\n
&&I|hw>=2\omega|hw>,~~~ N|hw>=2\lambda|hw>,\n
&&E_{13}|hw>=E_{32}|hw>=E_{24}|hw>=E_{14}|hw>=E_{12}|hw>=E_{34}|hw>=0.\n
\eeqa
Then state vectors are mapped into functions
\beqa
\psi_{J_1,J_2,\omega,\lambda}&=&<hw|e^{\a_{13}^\dg E_{13}+\a_{32}^\dg E_{32}
    +\a_{24}^\dg E_{24}+\a_{14}^\dg E_{14}+a_{12}^\dg E_{12}
    +a_{34}^\dg E_{34}}|\psi>|0>\n
&=&<hw|e^{\a_{13}^\dg E_{13}}e^{\a_{32}^\dg E_{32}}e^{\a_{24}^\dg
    E_{24}}e^{(a_{12}^\dg+\frac{1}{2}\a_{13}^\dg\a_{32}^\dg)E_{12}}\n
& &\times e^{(a_{34}^\dg+\frac{1}{2}\a_{32}^\dg\a_{24}^\dg)E_{34}}
   e^{(\a_{14}^\dg+\frac{1}{2}a_{12}^\dg\a_{24}^\dg-\frac{1}{2}
   \a_{13}^\dg a_{34}^\dg+\frac{1}{3}\a_{13}^\dg\a_{32}^\dg\a_{24}^\dg)
   E_{14}}|\psi>|0>,\n
\eeqa 
where the Zassenhaus formula (\ref{Zassenhaus}) has been used
repeatedly, and operators $A$ are mapped as before:
\beqa
A|\psi>\rtarr \G(A)\psi_{J_1,J_2,\omega,\lambda}&=&
   <hw|e^{\a_{13}^\dg E_{13}}e^{\a_{32}^\dg
       E_{32}}e^{\a_{24}^\dg E_{24}}\n
& &\times e^{(a_{12}^\dg+\frac{1}{2}\a_{13}^\dg\a_{32}^\dg)E_{12}}
    e^{(a_{34}^\dg+\frac{1}{2}\a_{32}^\dg\a_{24}^\dg)E_{34}}\n
& &\times e^{(\a_{14}^\dg+\frac{1}{2}a_{12}^\dg\a_{24}^\dg-\frac{1}{2}
   \a_{13}^\dg a_{34}^\dg+\frac{1}{3}\a_{13}^\dg\a_{32}^\dg\a_{24}^\dg)
      E_{14}}A|\psi>|0>.\n
\eeqa
Here $\a_{ij}^\dg~(\a_{ij})$ are fermion operators with number operators
$N_{\a_{ij}}$ and $a_{ij}^\dg~(a_{ij})$ are boson operators with number
operators $N_{a_{ij}}$. They obey relations
\beqa
&&\{\a_{ij},\a_{kl}^\dg\}=\delta_{ik}\delta_{jl},~~~~\lt(\a_{ij}\rt)^2=
   \lt(\a_{ij}^\dg\rt)^2=0,\n
&&[N_{\a_{ij}},\a_{kl}]=-\delta_{ik}\delta_{jl}\a_{kl},~~~~
  [N_{\a_{ij}},\a_{kl}^\dg]=\delta_{ik}\delta_{jl}\a_{kl}^\dg,\n
&&[a_{ij},a_{kl}^\dg]=\delta_{ik}\delta_{jl},\n
&&[N_{a_{ij}},a_{kl}]=-\delta_{ik}\delta_{jl}a_{kl},~~~~
  [N_{a_{ij}},a_{kl}^\dg]=\delta_{ik}\delta_{jl}a_{kl}^\dg,
\eeqa
and all other (anti-)commutators vanish.
Moreover, $\a_{13}|0>=\a_{32}|0>=\a_{24}|0>=\a_{14}|0>=a_{12}|0>=a_{34}|0>=0$.

Taking $E_{13},E_{31}$ etc in turn and after long but straightforward
computations, we find 
\beqa
E_{13}&=&\a_{13}+\frac{1}{2}\a_{32}^\dg a_{12}-\frac{1}{2}\lt(a_{34}^\dg
  +\frac{1}{6}\a_{32}^\dg\a_{24}^\dg\rt)\a_{14},\n
E_{31}&=&(\omega+J_1+J_2)\a_{13}^\dg-a_{12}^\dg\a_{32}+\a_{14}^\dg a_{34}
  -\frac{1}{2}\a_{13}^\dg\lt(N_{\a_{32}}+N_{a_{12}}+N_{a_{34}}
  +N_{\a_{14}}\rt)\n
& &-\frac{1}{12}\a_{13}^\dg\a_{24}^\dg\lt(\a_{32}^\dg a_{34}
  +a_{12}^\dg\a_{14}\rt),\n
E_{32}&=&\a_{32}+\frac{1}{2}\a_{13}^\dg a_{12}+\frac{1}{2}\a_{24}^\dg
  \lt(a_{34}+\frac{1}{3}\a_{13}^\dg\a_{14}\rt),\n
E_{23}&=&(\omega-J_1+J_2)\a_{32}^\dg+a_{12}^\dg\a_{13}-a_{34}^\dg\a_{24}
  +\frac{1}{2}\a_{32}^\dg\lt(N_{\a_{13}}-N_{\a_{24}}+N_{a_{12}}-
  N_{a_{34}}\rt)\n
& &+\frac{1}{6}\a_{32}^\dg\lt(\a_{13}^\dg
  a_{34}^\dg+a_{12}^\dg\a_{24}^\dg\rt)\a_{14},\n
E_{24}&=&\a_{24}+\frac{1}{2}\a_{32}^\dg a_{34}+\frac{1}{2}\lt(
  a_{12}^\dg-\frac{1}{6}\a_{13}^\dg\a_{32}^\dg\rt)\a_{14},\n
E_{42}&=&(\omega-J_1-J_2)\a_{24}^\dg+a_{34}^\dg\a_{32}+\a_{14}^\dg a_{12}
  +\frac{1}{2}\a_{24}^\dg\lt(N_{\a_{32}}+N_{a_{12}}+N_{a_{34}}
  +N_{\a_{14}}\rt)\n
& &+\frac{1}{12}\a_{13}^\dg\a_{24}^\dg\lt(a_{34}^\dg\a_{14}
  -\a_{32}^\dg a_{12}\rt),\n
H_1&=&2J_1-2N_{a_{12}}-N_{\a_{13}}-N_{\a_{32}}+N_{\a_{24}}-N_{\a_{14}},\n
H_2&=&2J_2-2N_{a_{34}}+N_{\a_{13}}-N_{\a_{32}}-N_{\a_{24}}-N_{\a_{14}},\n
I&=&2\omega,\n
N&=&2\lambda-2\lt(N_{\a_{13}}-N_{\a_{32}}+N_{\a_{24}}+N_{\a_{14}}\rt),\n
E_{12}&=&a_{12}-\frac{1}{2}\a_{24}^\dg\a_{14},\n
E_{21}&=&2J_1a_{12}^\dg-(\omega+J_2)\a_{13}^\dg\a_{32}^\dg-a_{12}^\dg\lt(
  N_{\a_{13}}+N_{\a_{32}}+N_{a_{12}}\rt)\n
& &-\lt(\a_{14}^\dg-\frac{1}{2}\a_{13}^\dg a_{34}^\dg\rt)\a_{24}
  +\frac{1}{2}\lt(a_{12}^\dg+\frac{1}{3}\a_{13}^\dg\a_{32}^\dg\rt)N_{\a_{24}}\n
& &+\frac{1}{2}\a_{32}^\dg\lt(\a_{14}^\dg+\frac{1}{2}a_{12}^\dg\a_{24}^\dg\rt)
  a_{34}+\frac{1}{4}\a_{13}^\dg\a_{32}^\dg N_{a_{34}}\n
& &-\frac{1}{2}\lt(a_{12}^\dg-\frac{1}{6}\a_{13}^\dg\a_{32}^\dg\rt)N_{\a_{14}}
  -\frac{1}{4}\lt(a_{12}^\dg\a_{24}^\dg+\a_{13}^\dg a_{34}^\dg
  +\frac{1}{6}\a_{13}^\dg\a_{32}^\dg\a_{24}^\dg\rt)a_{12}^\dg\a_{14},\n
E_{34}&=&a_{34}+\frac{1}{2}\a_{13}^\dg\a_{14},\n
E_{43}&=&2J_2a_{34}^\dg-(\omega-J_1)\a_{32}^\dg\a_{24}^\dg-a_{34}^\dg\lt(
  N_{\a_{32}}+N_{\a_{24}}+N_{a_{34}}\rt)\n
& &+\lt(\a_{14}^\dg+\frac{1}{2}a_{12}^\dg\a_{24}^\dg\rt)\a_{13}
  +\frac{1}{2}\lt(a_{34}^\dg-\frac{1}{3}\a_{32}^\dg\a_{24}^\dg\rt)
  N_{\a_{13}}\n
& &-\frac{1}{2}\a_{32}^\dg\lt(\a_{14}^\dg-\frac{1}{2}\a_{13}^\dg
  a_{34}^\dg\rt)a_{12}-\frac{1}{4}\a_{32}^\dg\a_{24}^\dg N_{a_{12}}
  -\frac{1}{2}\lt(a_{34}^\dg+\frac{1}{6}\a_{32}^\dg\a_{24}^\dg\rt)
  N_{\a_{14}}\n
& &+\frac{1}{4}\lt(a_{12}^\dg\a_{24}^\dg+\a_{13}^\dg a_{34}^\dg
  -\frac{1}{6}\a_{13}^\dg\a_{32}^\dg\a_{24}^\dg\rt)a_{34}^\dg\a_{14},\n
E_{14}&=&\a_{14},\n
E_{41}&=&(\omega+J_1-J_2)\a_{14}^\dg+\frac{1}{2}
  (\omega+J_1+3J_2)a_{12}^\dg\a_{24}^\dg\n
& &  +\frac{1}{2}(-\omega+3J_1+J_2)\a_{13}^\dg a_{34}^\dg
  -\frac{1}{3}(J_1-J_2)\a_{13}^\dg\a_{32}^\dg\a_{24}^\dg\n
& &-a_{12}^\dg a_{34}^\dg\lt(\a_{32}-\frac{1}{6}\a_{13}^\dg\a_{24}^\dg
  \a_{14}\rt)-\lt(\a_{14}^\dg+\frac{1}{2}a_{12}^\dg\a_{24}^\dg\rt)
  N_{\a_{13}}\n
& &-\frac{1}{2}\lt(\a_{13}^\dg a_{34}^\dg+a_{12}^\dg\a_{24}^\dg\rt)
  N_{\a_{32}}+\lt(\a_{14}^\dg-\frac{1}{2}\a_{13}^\dg a_{34}^\dg\rt)
  N_{\a_{24}}\n
& &-\lt(\a_{14}^\dg+\frac{1}{2}a_{12}^\dg\a_{24}^\dg
  +\frac{1}{12}\a_{13}^\dg\a_{32}^\dg\a_{24}^\dg\rt)N_{a_{12}}\n
& &  +\lt(\a_{14}^\dg-\frac{1}{2}\a_{13}^\dg a_{34}^\dg
  +\frac{1}{12}\a_{13}^\dg\a_{32}^\dg\a_{24}^\dg\rt)N_{a_{34}}.
\eeqa
This gives a boson-fermion realization of 
$gl(2|2)$ in the non-standard basis. This result reconfirms the
differential operator realization found in \cite{Ding03b}. Here 
we have corrected  a few misprints appeared in the expressions for
$d_{E_{31}}, d_{E_{21}}, d_{E_{43}}$ and $d_{E_{41}}$ in the eq.(3.3) of 
\cite{Ding03b}.

From the above realization, we may deduce realizations of subalgebras 
$gl(2|1)$, $gl(1|2)$ and $gl(1|1)$ with less numbers of bosons and
fermions. For the subalgebra $gl(2|1)$, which is generated by 
$\{I,E_{13},E_{31},E_{32},E_{23},
E_{12},E_{21},H_1,N\}$, we choose $a_{34}=a_{34}^\dg=\a_{24}=\a_{24}^\dg
=\a_{14}=\a_{14}^\dg=0$ to obtain
\beqa
E_{13}&=&\a_{13}+\frac{1}{2}\a_{32}^\dg a_{12},\n
E_{31}&=&(\omega+J_1)\a_{13}^\dg-a_{12}^\dg\a_{32}
  -\frac{1}{2}\a_{13}^\dg\lt(N_{\a_{32}}+N_{a_{12}}\rt),\n
E_{32}&=&\a_{32}+\frac{1}{2}\a_{13}^\dg a_{12},\n
E_{23}&=&(\omega-J_1)\a_{32}^\dg+a_{12}^\dg\a_{13}
  +\frac{1}{2}\a_{32}^\dg\lt(N_{\a_{13}}+N_{a_{12}}\rt),\n
H_1&=&2J_1-2N_{a_{12}}-N_{\a_{13}}-N_{\a_{32}},\n
I&=&2\omega,\n
N&=&2\lambda-2\lt(N_{\a_{13}}-N_{\a_{32}}\rt),\n
E_{12}&=&a_{12},\n
E_{21}&=&2J_1a_{12}^\dg-\omega\a_{13}^\dg\a_{32}^\dg-a_{12}^\dg\lt(
  N_{\a_{13}}+N_{\a_{32}}+N_{a_{12}}\rt).
\eeqa
Similarly we may deduce a realization of the subalgebra $gl(1|2)$, which
is generated by $\{I,E_{32},E_{23},E_{24},E_{42},
E_{34},E_{43},H_2,N\}$, by choosing $a_{12}=a_{12}^\dg=\a_{13}=\a_{13}^\dg
=\a_{14}=\a_{14}^\dg=0$. Finally, the subalgebra $gl(1|1)$ is
generated by $\{e\equiv E_{13}+E_{24}, f\equiv E_{31}+E_{42}, I, N\}$. 
Its realization
is deduced by choosing $a_{12}=a_{12}^\dg=a_{34}=a_{34}^\dg=\a_{32}
=\a_{32}^\dg=\a_{14}=\a_{14}^\dg=0$. We have
\beq
e=\a_{13}+\a_{24},~~~f=2\omega(\a_{13}^\dg
  +\a_{24}^\dg),~~~
I=2\omega,~~~N=2\lambda-2(N_{\a_{13}}+N_{\a_{24}}).
\eeq
Writing $\a\equiv \a_{13}+\a_{24}$ (and thus 
$\a^\dg\equiv\a_{13}^\dg+\a_{24}^\dg$,
$N_\a\equiv N_{\a_{13}}+N_{\a_{24}}$), we see that $gl(1|1)$ is realized in
terms of one fermion $\a,\a^\dg$.

\section{Conclusions and Discussions}

In this article we have discussed the application of the super coherent
state method to the construction of free boson-fermion realizations and
representations of superalgebras. We have worked through a few examples
corresponding to lower rank superalgebras. The representations are
constructed out of particle states in the super-Fock spaces.

As the results show, the computations become very involved with the
increase of the rank of superalgebras concerned. Nevertheless, 
it is expected that the 
procedure can be extended to higher rank superalgebras so as to obtain free
boson-fermion realizations and explicitly construct representations of the
superalgebras with the help of the realizations. 
In the previous section we have obtained
the free boson-fermion realization of the rank 3 non-semisimple
superalgebra $gl(2|2)$ in the non-standard basis. 
This superalgebra has applications in disordered systems and integer
quantum Hall plateau transition \cite{Zir99}. 
It is interesting to explicitly construct its finite-dimensional
representations by the method described in this paper. This then will
enable the construction of primary fields of the $gl(2|2)$ non-unitary
CFT. This is a problem under investigation.

\vskip.3in

\no {\bf Acknowledgments:}

My interest in the coherent state construction arose partly from
Max Lohe's talk ``Vector coherent states and quantum affine algebras" given in
the Third University of Queensland Mathematical Physics Workshop 
in Coolangatta on October 2-4, 2002. I thank Max Lohe for sending me
the talk material. Financial support from
the Australian Research Council is also acknowledged. 

\vskip.3in

\no {\bf Note added in proof:} 

All finite-dimensional typical and atypical representations of $gl(2|2)$
and corresponding primary fields of the $gl(2|2)$ non-unitary CFT in the
standard basis have recently been constructed by us in \cite{Zha05a,Zha05b}.

\bebb{99}

\bbit{Roz92} 
L. Rozanski and H. Saleur, \npb {376} {1992} {461}.

\bbit{Isi94}
J. M. Isidro and A. V. Ramallo, \npb {414}{1994}{715}. 

\bbit{Flohr01}
M. Flohr, \textit{Int. J. Mod. Phys.} {\bf A18}, (2003) 4497.

\bbit{Efe83}
K. Efetov, \textit{Adv. Phys}. {\bf 32}, (1983) 53. 

\bbit{Ber95}
D. Bernard, preprint hep-th/9509137.

\bbit{Mud96}
C. Mudry, C. Chamon and X.-G. Wen, \npb {466} {1996} 383.

\bbit{Maa97}
Z. Maassarani and D. Serban, \npb {489} {1997} {603}.

\bbit{Bas00}
Z.S. Bassi and A. LeClair, \npb {578} {2000} {577}.

\bbit{Gur00}
S. Guruswamy, A. LeClair and A.W.W. Ludwig, \npb {583} {2000} {475}.

\bbit{Lud00}
A.W.W. Ludwig, preprint cond-mat/0012189.

\bbit{Bha01}
M. J. Bhaseen,  J.-S. Caux ,  I. I. Kogan  and  A. M. Tsvelik, 
\npb {618}{2001}{465}.

\bbit{Bal88}
A.B. Balantekin, H.A. Schmitt and B.R. Barrett, \jmp {29} {1988} {1634}.

\bbit{Bal89}
A.B. Balantekin, H.A. Schmitt and P. Halse, \jmp {30} {1989} {274}.

\bbit{Que03}
C. Quesne, \textit{Int. J. Theor. Phys.} {\bf 43}, (2004) 545.

\bbit{Bra93}
A.J. Bracken, M.D. Gould and I. Tsohantjis, \jmp {34} {1993} {1654}.

\bbit{Bow96}
P. Bowcock, R-L.K. Koktava and A. Taormina, \plb {388} {1996} {303}.

\bbit{Ras98}
J. Rasmussen, \npb {510} {1998} {688}.

\bbit{Sch77} 
M. Scheunert, W. Nahm and V. Rittenberg, \jmp {18} {1977} {155};
\jmp {18} {1977} {146}.

\bbit{Mar80} 
M. Marcu, \jmp {21} {1980} {1277}; \jmp {21} {1980} {1284}.

\bbit{Ding03}
X.M. Ding, M. D. Gould, C. J. Mewton and Y. Z. Zhang, 
 \textit{J. Phys}. {\bf A36}, (2003) 7649.

\bbit{Zha03}
Y.Z. Zhang, \textit{Phys. Lett.} {\bf A327}, (2004) 442.

\bbit{Ding03b}
X.M. Ding, M. D. Gould and Y. Z. Zhang,
\textit{Phys. Lett}. {\bf A318}, (2003) 354.

\bbit{Zir99}
M.R. Zirnbauer, preprint hep-th/9905054.

\bbit{Zha05a}
Y.Z. Zhang and M.D. Gould, \jmp {46} {2005} {013505} (19 pages).

\bbit{Zha05b}
Y.Z. Zhang, X. Liu and W.L. Yang, \npb {704} {2005} {510}.

\eebb

\end{document}